\documentclass[sigconf,nonacm,10pt,screen]{acmart}
\usepackage{Style}

\begin{document}


\title{AI Factories: It's time to rethink the Cloud-HPC divide
}


\author{Pedro Garcia Lopez}
\author{Daniel Barcelona Pons}
\affiliation{%
  \institution{URV}
  \city{Tarragona}
  \country{Spain}
}

\author{Marcin Copik}
\author{Torsten Hoefler}
\affiliation{%
  \institution{ETH}
  \city{Zurich}
  \country{Switzerland}
}

\author{Eduardo Quiñones}
\affiliation{%
  \institution{BSC}
  \city{Barcelona}
  \country{Spain}
}

\author{Maciej Malawski}
\affiliation{%
  \institution{Sano \& AGH}
  \city{Krakow}
  \country{Poland}
}

\author{Peter Pietztuch}
\affiliation{%
  \institution{ICL}
  \city{London}
  \country{UK}
}

\author{Alberto Marti}
\affiliation{%
  \institution{OpenNebula}
  \city{Madrid}
  \country{Spain}
}

\author{Thomas Ohlson Timoudas}
\affiliation{%
  \institution{RISE}
  \city{Kista}
  \country{Sweden}
}

\author{Aleksander Slominski}
\affiliation{%
  \institution{IBM Research}
  \city{New York}
  \country{USA}
}

\begin{abstract}

The strategic importance of artificial intelligence is driving a global push toward Sovereign AI initiatives. Nationwide governments are increasingly developing dedicated infrastructures, called AI Factories (AIF), to achieve technological autonomy and secure the resources necessary to sustain robust local digital ecosystems.

In Europe, the EuroHPC Joint Undertaking is investing hundreds of millions of euros into several AI Factories, built atop existing high-performance computing (HPC) supercomputers. However, while HPC systems excel in raw performance, they are not inherently designed for usability, accessibility, or serving as public-facing platforms for AI services such as inference or agentic applications. In contrast, AI practitioners are accustomed to cloud-native technologies like Kubernetes and object storage, tools that are often difficult to integrate within traditional HPC environments.

This article advocates for a dual-stack approach within supercomputers: integrating both HPC and cloud-native technologies. Our goal is to bridge the divide between HPC and cloud computing by combining high performance and hardware acceleration with ease of use and service-oriented front-ends. This convergence allows each paradigm to amplify the other. To this end, we will study the cloud challenges of HPC (Serverless HPC) and the HPC challenges of cloud technologies (High-performance Cloud).

\end{abstract}




\maketitle


\section{Introduction}

Sovereign AI efforts are gaining momentum worldwide as countries recognize the strategic importance of controlling their own artificial intelligence capabilities. In response to growing concerns over technological dependence and data sovereignty, governments are investing in national AI infrastructures known as AI Factories (AIFs). These facilities are designed to ensure local control over critical AI resources, including data, compute, and models, while fostering innovation and economic competitiveness. In this direction, countries such as Japan, India, and Norway~\cite{aifs} have recently built AIFs using the NVIDIA AI Factory infrastructure solution.

The European Commission is investing € 1.5 billion in 13 AIFs in Finland, Germany (2x), Greece, Italy, Luxembourg, Spain, Sweden, Austria, Bulgaria, France, Poland, and Slovenia with plans to invest € 20 billion in creating up to 5 AI GigaFactories in the near future~\cite{aifactories}. The EU aims to become a major player in the AI race by heavily investing in the European supercomputing network to create powerful AI infrastructures and counter the lack of large European cloud platforms and AI services. Although promising, this can be a risky bet~\cite{aifdoubts} if this massive investment fails to create the right ecosystem for AI startups.

A significant challenge for European AIFs is to build on existing HPC supercomputers by upgrading or extending their current infrastructure. These clusters are currently designed for high-performance computing (HPC) experts, requiring specialized knowledge to provision and configure computing resources. HPC users are parallel programming experts who design efficient and scalable systems using low-level APIs such as MPI, OpenMP, CUDA, and OpenCL, among others. Unlike cloud providers that offer user-friendly services like Software as a Service (SaaS) or Platform as a Service (PaaS), supercomputers typically provide direct access to hardware nodes using tools like SLURM and GPFS. This lack of multi-tenancy, isolation, and easy-to-use services makes them less accessible to non-experts. At the same time, training Python data analysts or AI practitioners to use such HPC technologies is not a realistic solution.

AI Factories present a unique opportunity to combine the strengths of both HPC and cloud computing. HPC excels at large-scale hardware acceleration needed to train large language models and other generative AI systems, while cloud technologies provide scalable, elastic, and user-friendly environments ideal for inference workloads and agentic platforms. Their convergence could power a new generation of hyper-distributed AI applications, made possible by emerging paradigms such as Acceleration-as-a-Service (XaaS) \cite{xaas}.

In this article, we advocate for a dual stack in supercomputers, integrating both HPC and cloud technologies. We justify why HPC technologies alone cannot sustain AI workloads without cloud-native stacks. AIFs require two stacks, one with dedicated hardware and HPC technologies (SLURM, GPFS), and a second one with virtualized resources and elastic cloud technologies (Object Storage, Kubernetes).

However, we also claim that both stacks cannot be isolated from each other. We propose integration at several levels:

\begin{itemize}
\item The resource pools (especially GPUs) cannot be isolated in different clusters using HPC or cloud stacks. These expensive resources may be used from one scheduler or the other to avoid resource wastage.
\item The data infrastructure must be shared across both stacks. Components such as the data catalog, datasets, and data pipelines may operate on either stack. To support this, high-performance parallel data access must be ensured for both HPC systems (through a PFS) and cloud services (via Object Storage).
\item The compute infrastructure should deliver \sloppy high-performance containers with optimized dependencies and libraries for both stacks. We advocate for performance-portable containers that enable acceleration as a service across both HPC and cloud environments.
\item The AI infrastructure should be cross-cutting, providing a unified model catalog, SaaS interfaces with automated provisioning, and user-friendly AI pipelines. These pipelines should seamlessly integrate with the shared data infrastructure and be accessible from both stacks.
\item Finally, we believe that user-facing AI pipelines, such as inference and agent-based workloads, require AIF integration with external cloud resources. To support this, we advocate for the inclusion of cloud federation endpoints within AIFs, enabling seamless interoperability with external cloud providers and sovereign cloud initiatives.

\end{itemize}

\section{Comparing Cloud and HPC settings}

\begin{figure}[t]
    \centering
    \includegraphics[]{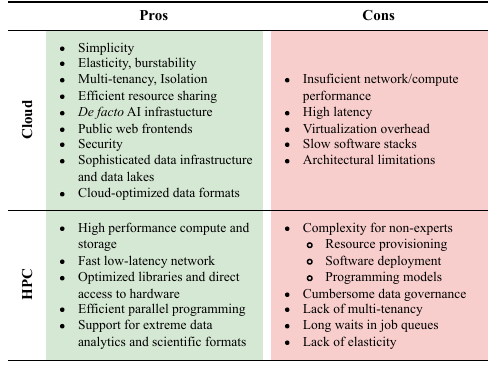}
    \caption{Comparison of Cloud and HPC}
    \label{fig:cloud_hpc}
\end{figure}

Cloud and HPC technologies are designed and evolved with different objectives and target applications in mind.
This has created stark differences in how these technologies are used and the performance that applications can squeeze from them.
In turn, both technologies have come with their particular pros and cons, which we review next (summarized in \cref{fig:cloud_hpc}).

HPC is offering its users direct access to dedicated hardware resources in order to optimize their computing jobs.
This means that users are responsible for provisioning resources and configuring software.
They are also expected to be experts in parallel programming, capable of scaling and optimizing their algorithms on the supercomputer.
In principle, HPC is ideally suited for long-running jobs that can take hours or even days, which may fit perfectly with the AI training of Large Language Models (LLMs), for example.

Although supercomputers are typically designed with the assumption that users are HPC and parallel programming experts, this is not always the case. Many non-expert users also utilize these powerful resources. In particular, in communities such as genomics, climate modeling~\cite{pangeo}, and physics, data scientists often rely on toolkits and frameworks such as PyTorch, Dask, and Spark to manage and process their data. These tools provide higher-level abstractions that simplify parallel computing and data handling, making supercomputers more accessible to those without deep expertise in parallel programming. This highlights the need for supercomputers to support a broader range of users and use cases, facilitating easier access and usability for both experts and non-experts alike.

Furthermore, giving non-expert users direct access to hardware and burdening them with resource management can create provisioning problems. First of all, there is a considerable risk of overprovisioning if they are not computing experts aware of the underlying infrastructure. An even more severe problem is the misuse of resources by non-parallelized applications (both compute and storage), as a result of using higher-level languages and toolkits such as Python. To overcome this problem, several wrapper technologies have been created to optimize general-purpose software libraries for HPC settings (such as Dask and Pytorch). However, since they rely on direct access to nodes using a job queue (Slurm), they cannot provide isolation or multi-tenancy, and they do not entirely prevent resource overprovisioning.

HPC offers low-latency communication and fast parallel storage compared to slower cloud storage.
However, it cannot offer elasticity, fast scaling, or burstability.
Jobs are assigned a static set of resources, and it is not straightforward to accommodate them to varying requirements.
Moreover, job queue management systems can take minutes or days to provision resources.

In contrast, clouds are public infrastructures for end users, which means that
\begin{inparaenum}[(i)]
    \item they are specialized in simplifying the user experience,
    \item they optimize resource utilization to increase benefits, and
    \item they offer strong security to protect against external attacks.
\end{inparaenum}
However, to achieve this dynamism, they use diverse virtualization technologies and high-level communication protocols (usually HTTP-based) that sacrifice performance.

These differences can be attributed to the characteristics of workloads that have dominated each field since its inception.
Although the software stack differs in many aspects, both cloud and HPC are increasingly accommodating both interactive and non-interactive jobs: batch jobs that have to tolerate hours or days of scheduling delays versus SLA-driven tasks requiring immediate execution (Table~\ref{tab:job_comparison}).
While clouds have historically been associated with dynamic resource allocation for end-user applications with unpredictable workloads, HPC systems are increasingly incorporating time-sensitive applications through \emph{urgent HPC} scenarios like earthquake prediction and wildfire modeling that demand an immediate response
~\cite{10.1007/978-3-030-34356-9_40,10.1145/3394277.3401853}.
The new demands introduced by AI can blur the apparent differences between the two systems, making them appear more similar: machine learning training represents long-running, statically scheduled batch processes, while inference demonstrates dynamic, elastic resource management.
Despite apparent differences between the software stacks of applications, such as Fortran-based numerical applications typical in HPC and GPU-accelerated and distributed Python frameworks for LLM training, the underlying infrastructure is converging.
The emergence of high-performance serverless~\cite{copik2023rfaas,10579231} computing demonstrates the demand for handling interactive workloads efficiently in both cloud and HPC (Section~\ref{sec:ai-inference-faas}), and this need will become even more critical with a more heterogeneous architecture in agentic AI~\cite{besta2025affordableaiassistantsknowledge}.

In summary, HPC focuses on performance and requires users to have expertise in parallel programming and resource provisioning on dedicated hardware. Instead, cloud computing focuses on optimizing multi-tenant access to virtualized resources, offering simplicity through higher-level abstractions such as PaaS, FaaS, and SaaS~\cite{xaas}. Whereas software stacks in HPC are highly efficient and optimized for their hardware (MPI, OpenMP, Fortran/C++), cloud stacks (Kubernetes, Spark, Python) are considerably slower and based on open HTTP protocols.

The problem is that both souls must be reconciled if AIFs are to become public digital infrastructures that support interactive and non-interactive workloads across both computing paradigms.
Both platforms were originally designed for their target user communities with different performance and simplicity objectives. AIFs offer a great opportunity for both technologies to cross-pollinate and evolve together. Let us study the open challenges of becoming AI Factories.

\begin{table}[t]
\centering
\caption{Interactive and non-interactive workloads.}
\label{tab:job_comparison}
\begin{tabular}{l|l|l}
& \textbf{Interactive} & \textbf{Non-Interactive} \\
\hline
Duration & Seconds, minutes & Hours, days \\
\hline
Allocation & Sub-second to seconds & Minutes to hours \\
\hline
Platform & FaaS, CaaS, Edge & IaaS, batch systems \\
\hline
Resources & Variable & Static, large \\
\hline
Scheduling & Immediate execution & Queue scheduling \\
\hline
\end{tabular}
\end{table}

\section{Towards an integrated dual stack}

AI Factories (AIFs) built on top of HPC supercomputers are now facing significant design challenges. When AIFs are developed from scratch, the most practical approach is often to adopt NVIDIA’s hardware and software platforms for AI factories \cite{nvidiaaifs}, which are inherently designed with a cloud-native architecture. A notable example of this is India’s Shakti-Cloud platform~\cite{shakti}. However, we see many cases where AIFs are constructed by upgrading or extending existing HPC supercomputing infrastructure, like is the case for most European proposals. In such scenarios, we identify three main architectural strategies: the conservative approach, the isolated dual-stack model, and the integrated dual-stack model.

The conservative design aims to maintain the entire HPC stack and create a thin web front-end using technologies like OpenOnDemand~\cite{chalker2024open} or small clusters using OpenStack or OpenNebula that will present a simple web interface to the underlying cluster resource manager (e.g., Slurm). The major risk of this approach is that AI practitioners will face a steep learning curve to deploy their AI workloads. Furthermore, such an approach precludes effective deployment of inference jobs or agentic platforms that require an elastic and dynamic environment capable of adapting to heterogeneous workloads.

The second alternative involves maintaining two isolated stacks for HPC and cloud computing. In this scenario, new hardware would be dedicated to a cloud-based stack within the NVIDIA ecosystem—such as the NVIDIA AI Factory~\cite{nvidiaaifs}, Dell AI Factory with NVIDIA, or IBM Vela~\cite{vela}. These are sophisticated platforms that provide managed, cloud-native tools for AI training and inference. As a result, the supercomputing system would consist of two distinct hardware and software environments—HPC and cloud—that operate independently and without awareness of each other. This separation means the two resource pools would be unable to share or utilize each other’s idle capacity. Additionally, if public, user-facing services are to be delivered via the AI Factory stack, administrators may face new security challenges, particularly if they lack experience managing public endpoints and the associated threat landscape.

\begin{figure*}
    \centering
    \includegraphics[width=0.8\textwidth]{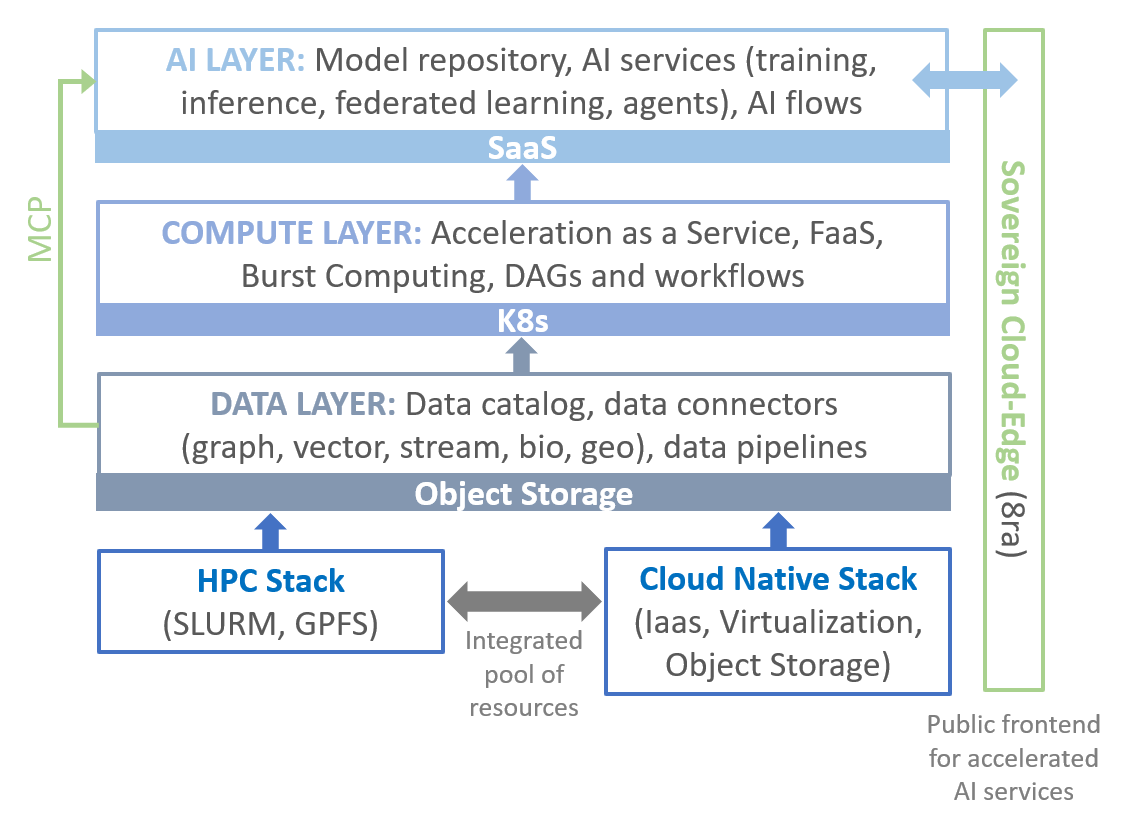}
    \caption{Integrated dual-stack architecture}
    \label{fig:dual}
\end{figure*}

We propose a third option: to build an integrated dual stack. As exemplified in \cref{fig:dual}, we still have two stacks: one for traditional, expert-driven HPC workloads and one for cloud-based, elastic workloads. In this model, all non-expert users are also moved to the cloud stack to benefit from higher-level SaaS services that simplify resource provisioning. The difference from the previous approach is that the schedulers of both stacks are integrated and may reuse idle resources from each other in a controlled way. The goal is to enable an efficient and shared usage of expensive GPU accelerators in both sides and to avoid isolated GPU islands.

The diagram illustrates that the data, compute, and AI components are built atop both the HPC and cloud stacks. It is evident that datasets must be accessible across both environments, and that advanced AI and compute services should be able to leverage resources from the system as a whole. On the right side of the figure, a vertical, cross-cutting layer connects the AI Factory with external cloud and edge resources. Crucially, public cloud services are envisioned as the primary front-end for delivering inference and agent-based services within the AI Factory. This requires some form of interconnection or federation with external public, private, and edge resources to support the deployment of advanced, distributed AI pipelines. Implementing an integrated dual-stack architecture brings significant challenges, particularly in data management, compute isolation and elasticity, and system accessibility.

In the next sections, we will explore the existing challenges to a dual stack integration in the data layer (Section 3.1), in the compute layer (Sections 3.2 and 3.3), and in the AI layer (Sections 3.4 and 3.5). Section 3.5 also explores how AI Factories will integrate with external sovereign cloud infrastructures to offer distributed AI pipelines.

\subsection{There is no AI without data}

The article titled similarly to this section~\cite{groger2021there} emphasizes that advances in AI depend fundamentally on data quality, availability, and management. 
Public cloud providers utilize Object Storage to build Data Lakes and Data Warehouses (e.g., AWS Data Exchange~\cite{awsexchange}, AWS Registry for Open Data~\cite{awsregistry}, Google BigQuery public datasets~\cite{googledatasets}, and Azure Open Datasets~\cite{azuredatasets}), offering massive-scale hosting, data catalogs for discovery, and tight integration with analytics services such as Google BigQuery and Amazon Athena. These infrastructures ensure that AI training pipelines can access, query, and govern datasets efficiently.

Object Storage has emerged as the \textit{de facto} universal solution in the cloud. Its immutable, object-centric design delivers practically unlimited scalability and bandwidth to data-intensive applications. The Amazon S3 API, widely adopted across platforms, promotes interoperability and application portability. Initially used for archival storage, now forms the backbone of cloud-native Big Data and AI stacks~\cite{levin2019aiops}, including Data Spaces, Data Lakes, and Data Warehouses.

In the HPC domain, research has shown that Object Storage can outscale traditional parallel file systems (PFS)~\cite{liu2018evaluation}, but platform-specific bottlenecks persist for certain HPC workloads. Other works have addressed S3 API limitations with middleware libraries~\cite{gadban2021analyzing, gadban2022analyzing} and MPI-aware connectors~\cite{araujo2023libcos}. Nevertheless, many applications, including AI training systems, still rely on high-throughput parallel file systems like Lustre or GPFS for concurrent, small-file access~\cite{10579231}. 

Although there is persistent concern among HPC administrators regarding performance, research has shown that these issues can be mitigated. Given that serving external users is crucial for AI systems, co-deployment of Object Storage and PFS is becoming the norm, balancing performance with multi-tenant governance and fine-grained access control.

Current supercomputing sites often lack integrated data platforms and open datasets. They typically function as compute-and-storage providers, with PFS that offer limited data sharing, governance, and (like Object Storage) small-file performance. Bridging this gap is essential to foster collaborative AI research and industrial use cases.

Four key data types in AI workflows introduce distinct challenges:
\begin{itemize}
    \item \textbf{Scientific formats:}
    Semi-structured domain-specific encodings require costly preprocessing---partitioning, filtering, or format conversion---to enable parallel processing. Example domains are genomics (FASTA, FASTQ, VCF, BAM), metabolomics (mzML, imzML), astronomy (MS), high-energy physics (ROOT) and geospatial data (LIDAR, GeoTIFF).

    \item \textbf{Vector embeddings:}
    Dense representations (from text or images) demand elastic embedding generation and scalable, adaptive vector databases to support key AI tasks such as retrieval-augmented generation (RAG).

    \item \textbf{Streams:}
    Real-time data feeds are key for conversational agents and autonomous systems. Achieving elastic stream processing and unifying batch/stream workflows remains an open problem (e.g., Lambda architecture\footnote{The Lambda architecture problem refers to the complexity of maintaining separate pipelines for batch and real-time processing.}~\cite{kiran2015lambda}). In practice, vector embeddings are often derived from data streams~\cite{finol2024streamsense} or scientific formats uploaded to Object Storage.

    \item \textbf{Graphs:}
    Used for modeling and reasoning on structured relationships in recommendation, geospatial, and knowledge-graph applications. Graph Neural Networks highlight the need for efficient graph storage and query interfaces~\cite{vatter2023evolution,10443519}.
\end{itemize}

To address these gaps, we argue that research and innovation should focus on:
\begin{itemize}
    \item Cloud-optimized formats and connectors for on-the-fly partitioning of scientific datasets (e.g., ZARR, COG, COPC). However, these formats require full conversion of legacy data, which involves heavy processing. Recent extensible data connector libraries~\cite{arjona2024dataplug} alleviate this problem and apply to both Object Storage and PFS.
    \item Serverless vector databases~\cite{vexless} tightly integrated with Object Storage, alongside commercial offerings (Pinecone, UpStash, Amazon OpenSearch).
    \item Streaming platforms with native tiering on Object Storage (Pravega~\cite{gracia2023pravega}, WrapStream~\cite{warp}), reducing dependency on disk-based brokers like Kafka.
    \item In-memory and hybrid graph databases~\cite{thimmaiah2025memory} that leverage Object Storage for scalable, long-term graph persistence.
\end{itemize}

\textbf{Cloud Challenge (Serverless HPC Data Platform)}:
AIFs should build serverless data infrastructures on Object Storage---providing open Data Catalogs, standard access protocols, and elastic query services---to host heterogeneous Data Hubs and efficiently process scientific formats, vector embeddings, real-time streams, and graphs. In the end, AI factories will compete to attract datasets in different vertical domains, as data are the real fuel for AI training and data analytics pipelines.

\textbf{HPC Challenge (Unified High Performance Storage)}:
Cloud tools should evolve to deliver high-bandwidth Object Storage tightly integrated with parallel file systems~\cite{gadban2021analyzing, gadban2022analyzing}---by optimizing S3-compatible APIs, MPI-aware connectors, and middleware---to create a unified, multi-tenant storage layer that satisfies both AI training pipelines and compute-intensive HPC workloads. Ultimately, the goal is a storage platform that supports identity management and data governance while delivering high aggregate bandwidth to parallel workloads and interactive services (streams, vectors, and graphs).

\begin{tcolorbox}[colframe=gray!80]
{\textbf{Data challenges}}

\textbf{Cloud:} AIFs should embrace Serverless Object Storage platforms with open catalogs, standard APIs, and elastic querying for diverse data types.

\textbf{HPC:} We need unified high-bandwidth Object Storage and parallel file systems via optimized S3 APIs and connectors for AI and HPC workloads.
\end{tcolorbox}

\subsection{AI training requires Kubernetes clusters and multi-tenancy}

Kubernetes (K8s) has become the de facto standard for deploying and managing AI workloads. In cloud environments, AI infrastructures are commonly built on large-scale Kubernetes clusters. For instance, OpenAI described scaling Kubernetes to over 7,000 nodes to train models like GPT-3~\cite{k8s7000}. IBM has also deployed massive Kubernetes environments designed for multi-tenant AI workloads~\cite{jayaram2019ffdl}, and the use of Kubernetes for Deep Learning is extensively documented in the literature~\cite{ye2024deep}. Most recently, IBM reported that its infrastructure for GenAI model development is built on Kubernetes and container technologies~\cite{gershon2024infrastructure}.

To attract AI startups and practitioners, supercomputing centers must support Kubernetes natively. This is essential for deploying modern AI frameworks and ecosystems. Additionally, other scalable data analytics technologies such as Dask and Apache Spark can be seamlessly deployed atop Kubernetes clusters.

Importantly, relying solely on wrapper tools that expose HPC resources (e.g., Slurm or GPFS) through Kubernetes-like abstractions does not offer a complete solution. Several studies have already explored the convergence of Kubernetes and HPC environments~\cite{zhou2021container, piras2019container, chazapis2023running, milroy2022one, zervas2022virtual, sochat2024flux}. For example, \citet{chazapis2023running} proposed a "High Performance Kubernetes" approach that translates Kubernetes container requests into Slurm jobs using container runtimes like Singularity or Apptainer. While such solutions simplify job submission, they do not fully replicate the elasticity, resource sharing, and workload orchestration capabilities found in native Kubernetes environments.

A fundamental limitation of these wrappers is that they typically rely on static, user-dedicated resource allocation via traditional batch queuing systems (e.g., Slurm). In contrast, public cloud providers optimize infrastructure usage by dynamically placing containerized workloads across virtualized environments, enabling stronger isolation and more efficient resource utilization.

\textbf{Cloud Challenge (Serverless Kubernetes Clusters)}: One major challenge is offering AI practitioners access to large, multi-tenant Kubernetes clusters with minimal operational overhead—especially when using high-value resources like GPUs. Public cloud providers already deliver such capabilities through fully managed services like AWS Fargate, IBM Code Engine, and Azure Container Apps. Supercomputing facilities could adopt similar strategies using emerging cloud-native AI platforms such as NVIDIA AI Factory, Dell AI Factory with NVIDIA, or IBM Vela. These platforms offer GPU-backed managed Kubernetes environments suitable for AI training.

However, adopting these solutions in isolation would create two parallel and disconnected infrastructures—HPC and cloud—each with separate GPU pools and orchestration layers. Given the high cost and limited availability of GPUs, this fragmentation would lead to inefficient resource utilization. A more effective approach would be to integrate both stacks seamlessly, enabling bidirectional resource provisioning. For example, Kubernetes workloads could be scheduled over Slurm-managed HPC nodes or over native cloud resources transparently, depending on workload requirements and resource availability.

\textbf{HPC Challenge (High Performance Kubernetes)}: Although Kubernetes is now standard for AI workflows, it still presents significant challenges when deployed in HPC environments. These include issues related to scaling, network performance, and multi-tenant GPU sharing. To address this, Kubernetes technologies must be adapted to leverage low-latency interconnects and deliver consistent performance under mixed workloads.

Public cloud providers and AI companies like OpenAI have already demonstrated expertise in managing large-scale Kubernetes infrastructure. HPC operators should similarly develop the expertise to adapt Kubernetes to their unique hardware, network configurations, and user needs. This includes optimizing access to shared GPU accelerators and ensuring fairness, isolation, and high throughput in multi-tenant scenarios.

NVIDIA’s AI Factory, based on its DGX SuperPOD architecture, currently represents the most advanced commercial solution for high-performance Slurm and Kubernetes at scale. However, reliance on proprietary technologies raises concerns about sovereignty and vendor lock-in. For sovereign AI initiatives—particularly in Europe and other regions prioritizing digital autonomy—there is a growing need to adopt open-source alternatives. Notable efforts in this direction include GPU resource management via OpenStack, OpenNebula, and IBM’s open-source Vela project.

\begin{tcolorbox}[colframe=gray!80]
{\textbf{Compute challenges}}

\textbf{Cloud:} AIFs should offer K8s deployments that leverage expensive GPU resources over an integrated pool of resources including  both Slurm-managed and cloud-native nodes.

\textbf{HPC:} To ensure digital sovereignty, AIFs must embrace open source solutions providing Acceleration as a Service, and High Performance K8s services for AI practitioners.

\end{tcolorbox}

\subsection{AI inference and data processing require FaaS and Burstability}
\label{sec:ai-inference-faas}

Serverless computing has redefined elasticity in the cloud by enabling fine-grained resource management, rapid scaling, low-latency execution, and cost-efficient pay-as-you-go billing. This model is ideal for dynamic workloads requiring fast scale-up, short execution times, and fluctuating resource needs --- such as real-time processing, event-driven analytics, and microservices. By abstracting infrastructure management, serverless computing allows developers to focus on application logic, fostering innovation with minimal operational burden.

Serverless data analytics has seen substantial growth in recent years~\cite{jonas2017occupy,fouladi2017encoding,zhang2021caerus,sampe2021outsourcing}, capitalizing on this elasticity to process large datasets efficiently. Similarly, serverless models have proven effective for AI inference~\cite{aslani2025machine, christidis2020enabling, ali2022optimizing}.

In HPC environments, early efforts have introduced FaaS systems like rFaaS~\cite{copik2023rfaas,10579231}, HPCWhisk~\cite{przybylski2022using}, funcX~\cite{li2022funcx}, and LithopsHPC~\cite{lithopshpc}. However, the rigidity of Slurm-based resource queues prevents true elasticity, as startup times can span minutes to hours—far from the instant scalability expected in serverless platforms.

To bridge this gap, next-generation supercomputers must support rapid burst allocation of compute resources --- launching hundreds of CPUs in milliseconds --- to emulate the responsiveness of platforms like AWS Lambda or IBM Code Engine. Such agility could unlock novel paradigms like Granular Computing~\cite{lee2019granular}.

The primary limitation of Slurm-like resource management lies in its design for long-term reservations (hours to days), whereas cloud FaaS platforms operate on short-lived, on-demand models using dynamic resource pools. Interestingly, the high-speed interconnects in HPC systems (e.g., InfiniBand, RDMA) can support low-latency RPCs~\cite{lee2019granular}, enabling superior real-time performance compared to cloud networks.
However, the software stack in the cloud does not cooperate well with InfiniBand and instead requires socket-based programming with IP addressing, which motivates the different interconnect technologies present in HPC and cloud to converge and create a unified ecosystem~\cite{9810039}.
This aligns with work on ultra-fast computation like Millisort~\cite{li2021millisort}, and emerging models like Burst Computing~\cite{barcelonapons2025burst} and FMI~\cite{10.1145/3577193.3593718}, which extend beyond FaaS to support elastic, dynamic MPI workloads.

\textbf{Cloud Challenge (Serverless HPC FaaS for AI Inference)}:  FaaS is particularly beneficial in two AI workflow stages: data preprocessing (ETL, streams) and inference. Supercomputers must support elastic resource provisioning for such short-lived tasks by integrating serverless technologies into both cloud and HPC stacks. Prior work~\cite{copik2023rfaas,przybylski2022using,li2022funcx,lithopshpc} has set a foundation, but deployment over elastic resource pools—drawing from idle HPC nodes~\cite{10579231} or available cloud capacity—is essential.

A critical need is the development of serverless data processing libraries for complex data formats, including streams, graphs, embeddings, and scientific formats. These libraries would support AI and data platforms like PyTorch, Ray, and Dask by offering scalable, on-demand ETL capabilities (see \cref{fig:dual}).

\textbf{HPC Challenge (High Performance FaaS)}: Cloud-based FaaS platforms, such as AWS Lambda, suffer from limitations: cold starts in the hundreds of milliseconds, limited cores per function, no GPU access, and constrained bandwidth (e.g., ~100 MB/s to S3). In contrast, HPC FaaS systems like rFaaS and LithopsHPC have demonstrated significantly reduced cold starts, inter-function communication, and superior compute and storage throughput.

Moreover, while public cloud FaaS often caps concurrency (e.g., 1,000–10,000 functions), HPC environments could scale further, supporting larger parallel workloads. This opens the door to dynamic, elastic computing models that combine MPI/OpenMP with FaaS~\cite{barcelonapons2025burst,li2021millisort}.

Finally, to support cross-domain AI inference and agentic platforms, integration across HPC, AIFs, and cloud stacks is crucial. If cloud-native platforms become the entry point to supercomputing services, AIFs must interoperate seamlessly with external clouds. Open-source orchestration frameworks like OpenStack and OpenNebula may serve as essential glue technologies, enabling unified scheduling, data sharing (e.g., via Object Storage and streams), and cross-stack orchestration between Kubernetes and FaaS platforms.

\begin{tcolorbox}[colframe=gray!80]
{\textbf{Burstability challenges}}

\textbf{Cloud:} AIFs should provide burstable serverless FaaS models to cope with stateless AI workloads in both HPC and cloud stacks.

\textbf{HPC:} AIFs should offer High Performance FaaS deployments with super-fast direct communication, and access to compute and network accelerators to solve complex elastic and stateful AI challenges.
\end{tcolorbox}

\subsection{Empowering AI practitioners with guardrails and SaaS}

Current supercomputers are not yet ready for the general public because they are restricted to IaaS-like models that require a strong technical background and resource provisioning skills. If supercomputers are going to open their infrastructure to external users, they should improve the user experience (UX) by allowing both PaaS/FaaS and SaaS services designed based on the realities of data science and AI engineering work.

Training an AI model and deploying it (for AI inference) is a highly iterative and experimental process involving many computationally demanding steps beyond just training algorithms. An empirical study of production ML pipelines at Google challenges the conventional wisdom that training algorithms are the most computationally costly part of ML development \cite{Xin2021Empirical}. This study found that data/model analysis and validation together accounted for a greater share of their compute costs than the actual training of models. It also found that data ingestion and data preprocessing together accounted for a similar share of their compute costs.

Activities such as data analysis and exploration, and model verification in production (during inference) are vital stages in the AI development process, and require collaboration between multiple teams and specialists. Many of these specialists may not even have traditional engineering backgrounds. A study of ML engineering in practice suggests that experimentation with data features may be more common than experimentation with different models, highlighting the need for tools to support such activities \cite{Shreya2024Production}. Results and insights from one experiment guides future experiments, and it is critical to increase experimental and developer velocity. This requires on-demand access to high-burst compute is therefore critical to the AI development process.

In practice, it is common that AI models are not trained from scratch. Instead, AI practitioners often experiment with different pre-trained models, configurations, and data transformations. This experimentation is often done in an interactive way with Jupyter notebooks, often adding new code along the way---and idling expensive compute resources unless they are shared.

Cloud providers have extensive experience in offering advanced SaaS services to users, including Google CoLab or Amazon SageMaker. They offer Python notebooks that considerably simplify the execution of AI data processing over cloud hardware. The most challenging aspect is providing Python notebooks that support interactive analytics for domains that handle large amounts of data.

For example, in HPC there are also JupyterHub or Jupyter notebook interfaces to Slurm resources, but in the end they are a simple façade to a batch platform so they cannot provision elastic or serverless resources that adapt to the current demand or data volumes. On the contrary, both Amazon and Google are offering elastic managed versions of Spark like Amazon EMR or Google Dataproc. Other online services like AnyScale Unified AI platform, Coiled.io, Modal.com and PyRun.cloud also claim automated elasticity of cloud resources with easy-to-use SaaS interfaces for AI practitioners.

Dataflow-based architectures are common in practice, using directed acyclic graphs (DAGs) to compose AI workflows combining various data processing, training, and deployment components, using workflow orchestrators such as TensorFlow \cite{TensorFlow}, Kubeflow \cite{Kubeflow}, Metaflow \cite{Metaflow}, and Apache Airflow \cite{ApacheAirflow}. Many of them are designed for cloud environments and to simplify the life of developers, but introduce inefficiencies such as unnecessary data copying, difficulties scaling caching strategies for big data, and wasted computations when one stage of a pipeline fails and intermediate results that could be reused are lost or discarded. Smart orchestration of AI pipelines using scheduling and caching strategies can significantly speed up the execution of AI training workflows \cite{Xin2018Helix,Behrouz2022ReuseOptimizations,Tagliabue2024FaaSFurious,Kontaxakis2024HYPPO,Li2025CachingFramework}.

The resource needs of these data processing and training pipelines are complex to estimate a-priori and vary depending on the specific data transformations and model configurations. It is not feasible to manually plan capacity needs and provision resources for training runs. Many of the largest data-driven organisations have developed their own cloud-native platform solutions precisely to deal with these issues and offer elastic resource management \cite{Uber2025Kubernetes}, and adapting them to new requirements from large AI models that must be sharded across nodes \cite{ByteDance2023ScalesInference}.

As we can see, the configuration of large PyTorch, Spark, Dask, or Flink clusters also requires advanced skills that must be simplified to end users to avoid resource waste and inefficiencies. Putting non-expert users 'on rails' means offering such managed platforms that simplify and automate the execution of such technologies.

\textbf{Cloud Challenge (Serverless AI Front-ends and automated resource provisioning)}:  Supercomputers that are willing to open their infrastructure to external users must offer higher-level services like PaaS and SaaS to their users. To put them on rails and ensure that they do not waste resources, it is important to offer managed data analytics services (Jupyter Notebooks, Spark, Dask, PyTorch). Here, it is important to devise intelligent automated resource provisioning techniques based on data that can simplify the user experience. The goal is to prevent users from provisioning resources by offering them high-level visual interfaces instead. Supercomputers should also expose OpenAI compatible APIs (via e.g. vLLM~\cite{kwon2023efficient}) and MCP (Model Context Protocol) servers to LLM platforms enabling rich interactions with AI workflows and agentic platforms. Providing such a managed or serverless experience helps users avoid the complex task of provisioning resources. 

\textbf{HPC Challenge (High Performance SaaS for AI)}:
Existing cloud offerings such as AWS EMR, Google Dataproc, or Anyscale’s AI platform primarily leverage general-purpose cloud resources rather than the advanced accelerators available in AI Factories (AIFs). This highlights a clear need for next-generation SaaS platforms capable of interfacing with both cloud-native environments and high-performance infrastructures through an integrated dual-stack architecture.

These platforms should support the full AI lifecycle, including data ingestion, preprocessing, training, and inference—seamlessly integrating data pipelines with sophisticated AI workflows. In this context, Model Context Protocol (MCP) servers can play a pivotal role by providing structured context to agentic platforms and AI pipelines. This not only simplifies user interactions but also ensures high performance and efficient orchestration across both stacks.

\begin{tcolorbox}[colframe=gray!80]
{\textbf{UX challenges}}

\textbf{Cloud:} AIFs should empower AI users with easy-to-use SaaS front-ends leveraging Low-Code, No-Code and AI-assisted development of AI pipelines.

\textbf{HPC:} AIFs should provide a serverless experience to resources and accelerators, simplifying deployment and optimization of AI pipelines.

\end{tcolorbox}

\subsection{Beyond AI Factories: extending their reach to the Compute Continuum}

AI Factories (AIFs) face a strategic choice: they can either evolve into full-fledged public cloud providers, delivering secure, user-facing services across a broad range of applications, or specialize in \emph{acceleration as a service (XaaS)}~\cite{xaas} while relying on external cloud and edge infrastructures to provide scalable, secure front-end access. In Europe, where most supercomputers lack the resources and expertise to operate as public cloud providers, it is more practical—and aligned with current policy directions—to integrate them within an external Cloud-Edge ecosystem. AI Factory support for integration with cross-platform workflow execution engines, such as ColonyOS \cite{Kristiansson2024HPCCloud,Kristiansson2024ColonyOS}, could be used to execute AI workflows across HPC-cloud-edge environments.

In this context, the recent initiative to establish AIF Antennas~\cite{antennas} aims to create the necessary infrastructure to ensure secure, remote access to AIF resources, enabling seamless integration with distributed users and systems.

Crucially, if AIFs are to serve as pillars of Sovereign AI, they must be embedded within Sovereign Cloud infrastructures that uphold principles of digital autonomy and data governance. The EU’s 8ra initiative~\cite{8ra} exemplifies this direction, with over 120 partners across 12 member states collaborating to develop a secure, interoperable, and sovereign Multi-Provider Cloud-Edge Continuum. Additionally, Europe is investing in large-scale strategic initiatives under the IPCEI (Important Projects of Common European Interest) framework~\cite{ipcei}, such as virt8ra~\cite{virt8ra}, which is already federating multiple European cloud providers within a virtualized infrastructure.

In the next years, we will see how these sovereign cloud efforts will seamlessly combine with AIFs to extend their reach to distributed AI pipelines in the entire Cloud-Edge Continuum. As we scale AI infrastructure, we are increasingly reaching the limits of what individual locations and data center sites can host. Individual data centers are limited by the availability of power, cooling, and other fixed infrastructure. In response, large-scale AI infrastructure begins to federate multiple data center locations to aggregate compute capacity to satisfy the ever-growing demand for AI workloads.

When supporting AI workloads across multiple data centers that form the Compute Continuum, the AI software stack is greatly impacted. In addition to new resource allocation and planning strategies that decide how to allocate data and computation to different locations, AI software stacks must also implement approaches for \emph{distributed AI} that decompose AI training and inference computation across multiple locations in the Compute Continuum.

This introduces new challenges because traditional machine learning stacks (e.g., PyTorch or vLLM) have not been designed with these kinds of deployment model in mind. Similarly, the software infrastructure that manages the control plane (e.g., Kubernetes) also struggles to make resource allocation and scheduling decisions when faced with distribution over multiple, potentially disparate data center sites.

Although there are specialized distributed AI approaches, such as \emph{federated learning}, they focus on narrow use cases, for example, when protecting the privacy of training data. Other proposals, such as Apple's Private Cloud Compute~\cite{applepcc}, combine edge devices and cloud infrastructure to execute AI inference tasks seamlessly to provide user privacy.

Instead of these specialized solutions, we see an opportunity for new AI system designs that embrace some of the recent advances in distributed AI (e.g., DiLoCo~\cite{douillard2023diloco}) and provide feasible system implementations that enable users to deploy large-scale AI workloads across the entire Compute Continuum. The goal here would be to scale AI infrastructure by combining resources from the Compute Continuum, and increase efficiency and sustainability by intelligently deciding on the use of different compute resources at different sites.

\section{Conclusion}
\label{Conclusion}

AI factories represent a significant economic investment in Europe's pursuit of a sovereign digital infrastructure to foster AI innovation. This initiative involves multimillion-euro investments in various supercomputing centers across the continent. By providing access to advanced and costly hardware, this cutting-edge infrastructure will unlock new opportunities for innovation and technological AI leadership.

However, this large-scale investment introduces significant risks and design challenges for AI factory (AIF) architects. On one hand, AI practitioners are accustomed to cloud-native environments ---such as Kubernetes and object storage--- and typically work with the sophisticated data infrastructures provided by cloud platforms. On the other hand, AI inference and agentic platforms demand elastic, secure front-ends that seamlessly integrate cloud and edge resources. In contrast, HPC supercomputers are traditionally optimized for batch processing, use entirely different software stacks, and are not designed to support user-facing applications or interactive front-ends.

Our proposed solution is a sovereign open source integrated dual-stack architecture that combines the strengths of both HPC and cloud technologies. HPC supercomputers excel at high-performance computing and can provide acceleration with specialized hardware for data- and compute-intensive tasks such as AI model training and fine-tuning. Meanwhile, cloud computing is inherently suited for user-facing services, offering the scalability and simplicity required for AI inference and agent-based applications.

To ensure efficient and unified operation of AI Factories (AIFs), HPC and cloud resource pools ---particularly, expensive GPUs--- must not be siloed, but instead managed through interoperable scheduling to prevent resource underutilization. A shared data infrastructure is essential, enabling seamless access to data catalogs, datasets, and pipelines across both environments, supported by high-performance storage solutions like GPFS and object storage. Similarly, compute infrastructure should rely on performance-portable containers optimized for both stacks, facilitating acceleration-as-a-service. The AI layer must provide a unified model catalog and intuitive SaaS pipelines that span both ecosystems and integrate tightly with shared data services. Finally, user-facing workloads such as inference and agent-based tasks demand that AIFs incorporate cloud federation endpoints, enabling scalable, secure interoperability with external and sovereign cloud infrastructures.

We anticipate significant differences in the design and implementation approaches across various AIFs. If overly conservative or proprietary strategies are adopted, they risk falling short of the ambitious objectives set for these initiatives and could lead to the underutilization, or even waste, of substantial public investments.


\bibliographystyle{ACM-Reference-Format}
\bibliography{References}

\end{document}